\pdfoutput=1
\documentclass{aastex701}

\newcommand{\edt}{2026~ED$_{138}$}

\shorttitle{TESS Precovery of 2026~ED$_{138}$}
\shortauthors{Han et al.}

\begin{document}

\title{Seven Years of TESS Precovery Observations of the Dwarf-planet Candidate 2026~ED$_{138}$}

\author[0000-0002-7127-7643]{Te Han}
\altaffiliation{These authors contributed equally.}
\affiliation{MIT Kavli Institute for Astrophysics and Space Research, Massachusetts Institute of Technology, Cambridge, MA, USA}
\email{tehan@mit.edu}

\author[0000-0002-4371-3460]{Jeroen Audenaert}
\altaffiliation{These authors contributed equally.}
\affiliation{MIT Kavli Institute for Astrophysics and Space Research, Massachusetts Institute of Technology, Cambridge, MA, USA}
\email{jeroena@mit.edu}

\author[0000-0002-5788-9280]{Daniel Muthukrishna}
\affiliation{MIT Kavli Institute for Astrophysics and Space Research, Massachusetts Institute of Technology, Cambridge, MA, USA}
\affiliation{Center for Astrophysics $\vert$ Harvard \& Smithsonian, 60 Garden Street, Cambridge, 02138, MA, USA}
\email{danmuth@mit.edu}

\author[0000-0002-1139-4880]{Matthew Holman}
\affiliation{Center for Astrophysics $\vert$ Harvard \& Smithsonian, 60 Garden Street, Cambridge, 02138, MA, USA}
\email{mholman@cfa.harvard.edu}

\author[0000-0003-2058-6662]{George~R.~Ricker}
\affiliation{MIT Kavli Institute for Astrophysics and Space Research, Massachusetts Institute of Technology, Cambridge, MA, USA}
\email{grr@mit.edu}

\begin{abstract}
The trans-Neptunian object \edt\ was announced in 2026 October as the first dwarf-planet candidate in our solar system discovered by the Vera C.\ Rubin Observatory. The Transiting Exoplanet Survey Satellite (TESS) observed it in five month-long sectors between 2019 and 2026, four of them before its discovery. We subtract every Gaia DR3 star from these crowded full-frame images with the TESS-Gaia Light Curve (TGLC) package, then remove static residuals with a running median. Stacking the images along its predicted path detects \edt\ in all five sectors at 5--21$\sigma$, and at 34$\sigma$ combined. The TESS positions alone confirm the JPL orbit, and its brightness is consistent with the 0.08~mag fading expected from its increasing distance and changing phase angle.
\end{abstract}


\section{Introduction}

The trans-Neptunian object \edt\ was announced\footnote{\url{https://rubinobservatory.org/news/rubin-first-dwarf-planet-candidate}} on 2026 October 27 as the first dwarf-planet candidate found by the Vera C.\ Rubin Observatory \citep{Ivezic2019}, from Rubin detections and archival DECam images back to 2016 \citep{MPEC2026U85}. The orbit JPL computed from these 74 observations (solution JPL\#1\footnote{JPL Small-Body Database, \url{https://ssd.jpl.nasa.gov/tools/sbdb_lookup.html\#/?sstr=2026\%20ED138}}; hereafter the JPL orbit) has $a = 42.948 \pm 0.003$~au, $e = 0.14174 \pm 0.00006$ and $i = 24.4710 \pm 0.0001^\circ$, and its absolute magnitude, $H = 2.92 \pm 0.16$, places it among the brightest known trans-Neptunian objects. Its $V \approx 19.7$ and 600--1500~km size estimate suggest it escaped earlier surveys because it lies near the crowded Galactic plane.

The Transiting Exoplanet Survey Satellite \citep[TESS;][]{Ricker2015} takes full-frame images (FFIs) every 1800, 600 or 200~s and observed this field in five 27-day sectors between 2019 and 2026. Faint moving objects can be recovered from FFIs by shifting and stacking them along a predicted path \citep{RiceLaughlin2020, Nguyen2024}, but here 7--21 Gaia sources fall in every $21''$ TESS pixel. We remove them with the TESS-Gaia Light Curve package \citep[TGLC;][]{Han2023} and detect \edt\ in all five sectors.

\section{Data and Method}

We computed the TESS-centric ephemeris of \edt\ at each mid-exposure with JPL Horizons \citep{Giorgini1996}; its $3\sigma$ uncertainty is at most $0.5''$ (0.025 pixel). In all five sectors it fell on TESS Camera~1 and moved 74--90 pixels.

We used FFIs from the TESS Image CAlibrator \citep[TICA;][]{Fausnaugh2020}. In a $140\times140$-pixel region around the path in each sector, TGLC v0.8.1 models each image as the sum of all $(1.6$--$3.5)\times10^5$ Gaia DR3 sources \citep{Gaia2016, Gaia2023} at their catalog positions, with fluxes predicted from their Gaia photometry and an effective point-spread function (ePSF) fitted to the image, plus a planar background. Subtracting this model reduces the pixel-to-pixel scatter to 7--17\% of the original. The residuals are dominated by static, few-per-cent errors in the predicted stellar fluxes, which we remove by subtracting from each pixel its median over images taken 1.5--3~d earlier and later, masking pixels within 3 pixels of \edt\ (enclosing most of its ePSF) to avoid self-subtraction. This reduces the scatter to 0.5--1.7\%, 1.5--2.7 times below a running median applied without the TGLC model.

We measure the flux of \edt\ in each image by fitting the image's ePSF at its predicted position (forced PSF photometry). The images of a sector are combined with inverse-variance weights. An image's variance is the spread of the same photometry across 84 control tracks in that image (copies of the path offset 8--48 pixels sideways, so they never contain \edt), scaled by the pixel noise under the ePSF at the position of \edt. This scaling down-weights images in which \edt\ crosses bright or variable stars, without using its flux, and raises the significance 1.3--1.9 times. The uncertainty $\sigma$ of each sector's combined flux is the scatter of the same weighted combination over the control tracks. As a check of the flux scale, twenty copies of \edt\ injected per sector and measured the same way are recovered with 90--100\% of their flux, so we use the measured fluxes without correction.

\edt\ is too faint to locate in single images, so we measure its position from the stack: we shift the whole JPL path by a common offset on a grid (0.25-pixel steps, refined to 0.05 pixel), repeat the forced photometry on the star-subtracted images, and take the offset that maximizes the combined flux as the position at the flux-weighted mean time. We do this per half-sector when both halves exceed $5\sigma$ (Sectors 66, 93 and 104) and per sector for the other two, giving eight positions, with uncertainties from the larger of the scatter of eight injected copies and the flux-peak width divided by the S/N. We fit a two-body barycentric orbit, including light travel time and the position of TESS, to these positions alone.

\begin{figure}[ht!]
\centering
\includegraphics[width=\textwidth]{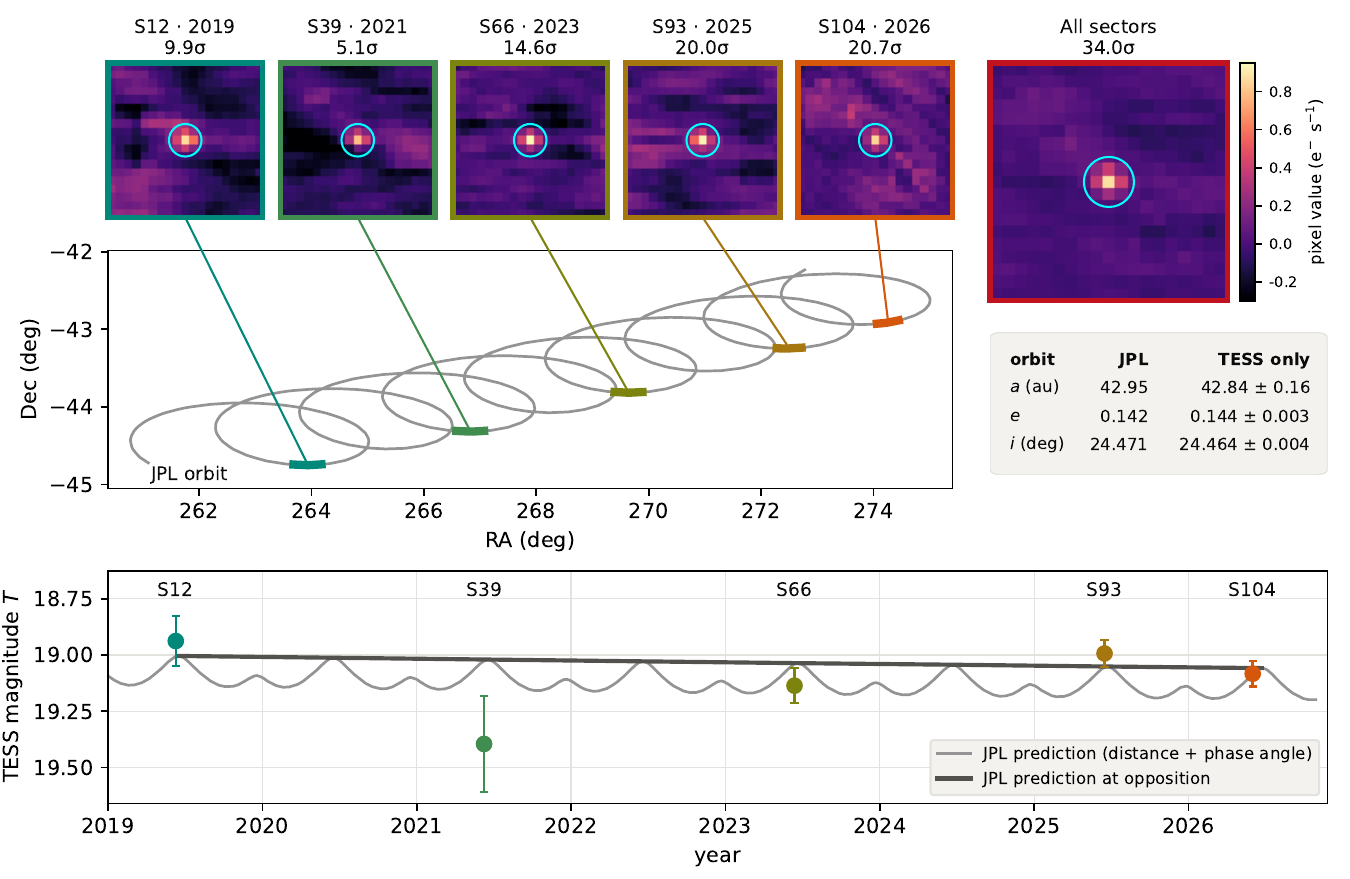}
\caption{Top: TESS images ($19\times19$~pixels, $6.5'$) stacked along the path of \edt\ after subtracting all Gaia DR3 stars and static residuals, per sector and all sectors combined (red), on one color scale, with the significance in units of the control-track scatter $\sigma$. Middle: the sky path of \edt\ seen from TESS (JPL orbit), each sector's segment, and the orbital elements from JPL and from TESS alone. Bottom: TESS magnitudes, with JPL's prediction (thin line) and its value at opposition, when TESS observes (thick line). Scattered light may explain the fainter magnitude and lower precision of Sector~39.}
\label{fig:stamps}
\end{figure}

\section{Results}

\textit{Detection.} \edt\ is detected at 9.9, 5.1, 14.6, 20.0 and 20.7$\sigma$ in Sectors 12, 39, 66, 93 and 104, and at 34$\sigma$ combined (Fig.~\ref{fig:stamps}). We applied three control tracks along the path of \edt\ at the wrong times of each sector (reversed in time, or shifted by $\pm4$~d), which would detect any static source on the path; all stay below $3\sigma$.

\textit{Brightness.} The TESS magnitudes are $T = 18.94 \pm 0.11$, $19.40 \pm 0.21$, $19.14 \pm 0.08$, $18.99 \pm 0.06$ and $19.08 \pm 0.06$. From 2019 to 2026 its distances from the Sun (46.5 to 47.2~au) and from TESS increased, dimming it by 0.063~mag, while the phase angle increased from 0.47 to 0.61$^\circ$, adding 0.017~mag ($H$--$G$ law, $G = 0.15$). The magnitudes agree with this 0.080~mag fading ($\chi^2 = 6.0$ for 4 degrees of freedom) and equally well with constant brightness ($\chi^2 = 6.1$); a linear trend fitted to the residuals from the prediction is $-0.009 \pm 0.016$~mag~yr$^{-1}$. Within sectors, half-day binned light curves scatter no more than those of the control tracks, and no period between 2~hr and 10~d stands out.

\textit{Color.} The offset that best fits the JPL $V$-band prediction to our $T$ magnitudes is $V - T = 0.65 \pm 0.03$ (plus systematic uncertainties of 0.16~mag from $H$ and 0.08~mag from the TGLC zero point), against the solar $V - T = 0.58$, from the solar $G_{\rm BP} - G_{\rm RP} = 0.82$ \citep{Casagrande2018} and the Gaia photometric relations \citep{Riello2021, Stassun2019}. The Rubin photometry gives $g - r = 0.64$ and $r - i = 0.25$ \citep{MPEC2026U85}, redder than the Sun, and its four $i$-band measurements during Sector~104 (mean $i = 19.41$) give a simultaneous $T - i = -0.33 \pm 0.07$.

\textit{Orbit.} The eight positions, with 1.2--4.8$''$ uncertainties, scatter about the JPL orbit as expected ($\chi^2 = 18.2$ for 16 measurements). Fitted alone, they give the heliocentric osculating elements $a = 42.84 \pm 0.16$~au, $e = 0.1435 \pm 0.0029$ and $i = 24.464 \pm 0.004^\circ$; the JPL orbit, 40--55 times more precise, is consistent with them ($\Delta\chi^2 = 6.1$ for six parameters). Parallax within each sector constrains the distance to $\pm0.06$~au.

\section{Summary}

TESS recorded \edt\ in five sectors over seven years, four of them before its discovery, in one of its most crowded fields. The same approach can extend TESS precovery of new Rubin discoveries into the Galactic plane.

\begin{acknowledgments}
This paper includes data collected by the TESS mission. Funding for the TESS mission is provided by NASA's Science Mission Directorate. This work has made use of data from the European Space Agency (ESA) mission Gaia, processed by the Gaia Data Processing and Analysis Consortium (DPAC). Funding for the DPAC has been provided by national institutions, in particular the institutions participating in the Gaia Multilateral Agreement. We thank the Rubin Observatory Solar System team for the discovery and astrometry of \edt. We used the AI assistants Claude (Anthropic) and Codex (OpenAI) to write code, make the figure and polish the text, which the authors reviewed and edited.
\end{acknowledgments}

\facilities{TESS, Gaia}

\software{TGLC \citep{Han2023}, Astropy \citep{Astropy2022}, NumPy, SciPy, Matplotlib}

\bibliography{ed138}{}
\bibliographystyle{aasjournalv7}

\end{document}